# Aberration-Free Imaging Based on Parity-Time Symmetric Nonlocal Metasurfaces

Francesco Monticone[1]*, Constantinos A. Valagiannopoulos[1,2]*, and Andrea Alù[1,3+]

[1]Department of Electrical and Computer Engineering, The University of Texas at Austin, Austin, TX 78712, USA

[2]Department of Radio Science and Engineering, Aalto University, Finland

[3]FOM Institute AMOLF, Amsterdam, The Netherlands

+alu@mail.utexas.edu

* These authors contributed equally to this work

*Lens design for focusing and imaging has been optimized through centuries of developments; however, conventional lenses, even in their most ideal realizations, still suffer from fundamental limitations, such as limits in resolution and the presence of optical aberrations, which are inherent to the laws of refraction. Although some of these limitations have been at least theoretically relaxed with the advent of metamaterials, several challenges still stand in the path toward ideal aberration-free imaging. Here, we show that the concept of parity-time symmetry, combined with tailored nonlocal response, allows overcoming some of these challenges, and we demonstrate the design of a loss-immune, linear, transversely invariant, aberration-free planarized metamaterial lens.*

# 1. INTRODUCTION

Imaging indicates the ability of accurately reproducing an ensemble of point sources from an "object space" into a different "image space". Although such regions of space are three-dimensional (3D), conventional imaging systems cannot reproduce full 3D images without distortion; instead, the imaging process is usually performed on pairs of two-dimensional (2D) transverse planes. For example, when we look around us, our eyes create 2D images on the retina, while the crystalline lens is adjusted to focus objects on planes at different distances. It is well known that optical imaging systems generally suffer from different kinds of *monochromatic aberrations* (spherical, comatic, astigmatic, etc.), which practically prevent the realization of an ideal image, even for a single object plane [1]. Therefore, it is relevant to ask whether ideal, aberration-free 2D or 3D optical imaging is at all theoretically possible, consistent with the laws of reflection and refraction (throughout this paper, ideal imaging is intended in the sense of geometrical optics, i.e., still limited by diffraction). Interestingly, it was demonstrated by Maxwell [2] (and later generalized) that no optical system *with a given focal length* can realize aberration-free imaging of two, or more, distinct transverse planes, no matter how complicated the lens (or combination of lenses) is made (see, e.g., [1],[3] and references therein). Instead, an ideal optical system that performs imaging of a 3D region of space is necessarily *afocal*. In such a device, known as an "absolute optical instrument", the optical length of any curve in the object space is preserved in the image space [1]. For over a century, it was thought that the only real example of a *planar* absolute optical instrument was a plane mirror [1],[4]-[5], which is, however, of limited interest, since the image points are virtual (Fig. 1a), i.e., they cannot be directly projected onto a screen without using an additional lens that would introduce aberrations.

Interestingly, with the advent of metamaterials it has been shown that a planar slab having a negative index of refraction can realize, at least in principle, a transversely-invariant, aberration-free, 'perfect' lens [7]-[8]. Such a double-negative (DNG) slab indeed acts as a planar absolute optical instrument, but – different from a mirror – it produces *real* images of the 3D object space. In addition, different from any other lens previously devised, a DNG lens is able to overcome Abbe's diffraction limit of conventional optical imaging systems, hence opening the possibility of realizing images with ideally unlimited transverse resolution [8]. Although this fact has generated large interest in the scientific communities, it was soon realized that such extreme imaging properties inevitably come at the expense of significantly increased sensitivity to losses, granularity, and other nonidealities, which are unavoidable in any practical implementation [9]-[10]. Besides, as confirmed by different numerical and experimental observations [10]-[11], the amplification of the evanescent spectrum, while supporting subdiffractive transverse resolution, inevitably determines strong resonant fields at the exit interface of the DNG lens, which makes it difficult to distinguish the images of different point sources along the longitudinal direction. In other words, in an ideal DNG lens the improvement in *transverse* resolution comes at the cost of reduced *longitudinal* resolution. All these issues have so far hindered the applicability of DNG lenses in practical scenarios, and several alternative solutions have been investigated to realize negatively refracting lenses without negative-index materials, based for example on phase conjugation, parametrically modulated or nonlinear surfaces [12]-[13], or higher-order diffraction in photonic crystal slabs [14]-[15]. While these solutions offer some advantages compared to DNG lenses, they are affected by other limitations that may fundamentally hamper their performance: phase conjugating lenses suffer from efficiency issues, as well as strong sensitivity to losses and other nonidealities, whereas photonic-crystal designs, not being transversely homogenous, are limited in the transverse resolution that they can achieve. Here, we explore a different route to imaging beyond the limitations of classical optics, based on parity-time (PT) symmetric systems that can realize loss-immune, transversely invariant, aberration-free lenses [16], having the imaging properties of absolute optical instruments, without some of the limitations of existing metamaterial designs.

## 2. ABERRATION-FREE IMAGING AND PARITY-TIME SYMMETRY

Consider again the case of optical imaging by a plane mirror, sketched in Fig. 1a. As mentioned above, an ideal mirror produces aberration-free *virtual* images of a *real* object space. Our goal is to devise an optical system that realizes similar imaging properties, but in which both object and image spaces are real. As a first step, imagine to go "through the looking glass" and assume that the virtual rays (red dashed arrows in Fig. 1a) beyond the mirror are real. In this scenario, we see that the imaging process is supported by a *backward phase and power flow* with respect to the positive $z$ axis. Interestingly, we show in the following that such phase and power flow distributions are actually attainable in a nonlocal PT-symmetric system, opening the possibility of realizing the equivalent of aberration-free imaging by a mirror, but within a real image space.

To make our discussion as general as possible, we start by considering the fundamental conditions to realize ideal imaging of a region of space, and we then derive – from first principles – the specific requirements that a structure should fulfill to implement this functionality. In analogy with the scattering response of a DNG lens, the scattering matrix of a generic two-port, linear, time-invariant system that performs ideal imaging should have, for any impinging propagating plane wave, the general form

$$S = \begin{pmatrix} 0 & e^{-i\beta d} \\ e^{-i\beta d} & S_{22} \end{pmatrix}, \tag{1}$$

which implies that (a) the structure should be reflectionless (i.e., impedance matched), at least from the input side, whereas, for now, we let $S_{22}$ be an arbitrary complex quantity; (b) the structure should "rewind" the phase of waves propagating through a distance $d$ with wavenumber $\beta = k_0 \cos\theta$, where $k_0$ is the free-space wavenumber and $\theta$ is the incidence angle [ we assume a time-harmonic convention $e^{-i\omega t}$ ]; (c) the appropriate phase advance should be realized for any angle of incidence, and for any source position, implying that the structure should be *transversely invariant*. It is also important to note that in Eq. (1) we assumed reciprocity, but not necessarily passivity, namely, the *S* matrix may be non-unitary.

In the special case $S_{22} = 0$ (which implies loss-free response), the scattering response in Eq. (1) can be implemented using an isotropic slab with refractive index $n = -1$, which supports impedance matching and negative phase velocity for any angle [7]-[8]. Given the challenges associated with realizing such ideal DNG metamaterial response, we explore whether other structures can implement Eq. (1) without requiring magnetic, or negative-index media. We consider a structure composed of two generic elements separated by a free-space region of length *d*, as sketched in Fig. 1b, a design appealing for several reasons, as it will become clear in the following. Using a transfer-matrix formalism applied to the considered structure [17], it is possible to derive the general requirements on the left and right elements to implement the desired *S* matrix. In particular, we perform an *eigendecomposition* of the transfer matrix *T* corresponding to Eq. (1) and identify the different blocks of this expansion as different elements of the structure. By further requiring that these elements are symmetric and reciprocal, we find that, in order for the system of Fig. 1b to realize ideal imaging, the transfer matrices of the left and right blocks need to have the general form

$$T_L = T_R^{-1} = \begin{pmatrix} \dfrac{1}{2c} & \dfrac{Z_0}{2}\left(\dfrac{1 \mp 2c}{c}\right) \\ \dfrac{1}{2Z_0}\left(\dfrac{1 \pm 2c}{c}\right) & \dfrac{1}{2c} \end{pmatrix}, \qquad (2)$$

where $Z_0$ is the wave impedance in free space, which depends on *incidence angle and polarization*, and $c$ is an arbitrary complex quantity. In addition, the requirements of symmetry and reciprocity impose $S_{22} = \mp\left(1 - e^{-2i\beta d}\right)/c$, which implies that the scattering matrix of the overall system, given by Eq. (1), is generally non-unitary, as energy gets absorbed and/or emitted by the structure.

As a relevant example, if we choose $c = 1/2$ and the upper sign in Eq. (2), the transfer matrices for the left and right elements of the system in Fig. 1b become

$$T_L = \begin{pmatrix} 1 & 0 \\ \dfrac{1}{Z_0/2} & 1 \end{pmatrix}, \quad T_R = \begin{pmatrix} 1 & 0 \\ -\dfrac{1}{Z_0/2} & 1 \end{pmatrix}. \qquad (3)$$

Given their symmetry, these transfer matrices can be realized by a pair of shunt-impedance sheets, or ultrathin metasurfaces, which corresponds to the structure originally considered in Ref. [16]. In particular, the left element of the pair is passive with positive resistance $R_L = +Z_0/2$, whereas the right element is active with negative resistance $R_R = -Z_0/2$. This system is composed of a *parity-time symmetric pair of metasurface elements*, i.e., it is invariant under reflection of time and space. By duality, choosing the opposite sign in Eq. (2) and $c = 1/2$, the same scattering response is obtained with a pair of *series-impedance sheets*. More in general, the *S* matrix (1) for ideal imaging can be realized with a pair of structures in free space having transfer matrices given by Eq. (2) – one acting as the time-reversed counterpart of the other – separated by a distance *d*. Note that the value of the parameter *c* and the sign choice in (2) only affect the value of $S_{22}$ in (1), whereas the phase advance and the unidirectional reflectionless response are ensured by the general form of the matrices in Eq. (2). These results confirm and generalize the findings in Ref. [16], showing that PT-symmetric metasurface pairs are ideal to realize aberration-free imaging.

## 3. NONLOCAL PT-SYMMETRIC METASURFACES

In a different context, the concept of PT-symmetry has recently sparked large attention in both quantum and classical physics [16]-[26]. Classical optics provides a particularly fertile ground for these concepts, as PT-symmetric systems can be readily realized with spatially balanced distributions of gain and loss. The unusual properties of PT-symmetric structures have been shown to enable loss compensation, exotic scattering effects, unidirectional invisibility, and many other interesting features that go beyond the limits of conventional passive systems [21]-[26]. In [16] it was realized that two ultrathin parallel metasurfaces designed to have impedance $R = \pm Z_0/2$ as in (3), *for a certain incidence angle and polarization*, the pair indeed becomes unidirectional reflectionless, and it supports negative phase and energy velocity, flowing backward from the active to the passive metasurface, implementing a scattering response consistent with Eq. (1) for the specific plane wave of interest. In order to use this feature to image a point source, in [16] a transversely inhomogeneous impedance profile of the two PT-symmetric surfaces was considered.

This scenario unfortunately works only for a specific object point – and it operates somewhat analogous to a hologram – with a response inevitably distorted if the source is moved or changed.

In order to achieve aberration-free imaging, on the contrary, an all-angle negative refraction response is required, implying that the condition $R = \pm Z_0/2$ on the surface impedance needs to be fulfilled *for any angle of incidence*. This is challenging because, as we mentioned above, the wave impedance of obliquely incident plane waves is naturally dispersive with the incidence angle $\theta$, i.e., $Z_0(\theta) = \eta_0/\cos(\theta)$ for TM polarization (magnetic field orthogonal to the incidence plane). A PT-symmetric metasurface pair can therefore achieve all-angle negative refraction only if the metasurface impedance becomes *spatially dispersive*, with a specifically prescribed angle dependence that compensates the natural angular dispersion of the wave impedance.

Based on these considerations, our goal is to design a PT-symmetric metasurface pair that implements this required nonlocal response. If the left element has transfer matrix exactly equal to Eq. (2) for any angle of incidence, its corresponding scattering matrix becomes

$$S_L = c \begin{pmatrix} \mp 1 & +1 \\ +1 & \mp 1 \end{pmatrix}, \qquad (4)$$

independent of the angle of incidence. Interestingly, as discussed in [17], this scattering matrix represents the response of an *omnidirectional coherent perfect absorber* (CPA), namely, a device that absorbs the complete angular spectrum of plane waves when illuminated at the same time from both sides with waves satisfying a given phase relation (which generalizes the recently proposed concept of coherent perfect absorber [27]-[28] to all-angle operation). Different from a conventional absorber (such as, e.g., [29]-[31]), a CPA only provides full absorption when it is bi-laterally illuminated, whereas its reflection and transmission are non-zero when excited from a single side. It can be shown [17] that, when a symmetric structure is bi-laterally illuminated with waves having equal amplitude and relative phase difference $\Delta\phi$, the omnidirectional CPA operation requires that the reflection $R$ and transmission $T$ coefficients respect the condition

$$R - T\,e^{i\Delta\phi} = 0 \quad \forall \theta\,, \tag{5}$$

where $R$ and $T$ do not depend on the angle of incidence.

It follows that the scattering matrix of the left element, given by Eq. (4), represents an omnidirectional CPA with $\Delta\phi = 0, \pi$, $R = \mp c$ and $T = c$, achieving perfect absorption for symmetric or anti-symmetric illumination, respectively.

In Fig. 1b we show a sketch of the complete "PT-symmetric lens", with the expected ray trajectories under point source illumination. As qualitatively seen in this ray picture, the lens functionality is indeed based on the fact that the left passive element operates as coherent perfect absorber, whereas the right active element, which is the time-reversed version of the CPA, acts as a lasing system for all excitation angles, amplifying the incoming signals and reproducing them symmetrically and coherently on both sides. We first focus on the design of the CPA, considering that the lasing element will be obtained by time-reversing the CPA response.

In an implementation based on zero-thickness metasurfaces, as in [16], the condition of angle independency of the scattering parameters in Eq. (4) indeed requires the impedance to become angle dependent (spatially dispersive), in order to compensate the angular dispersion of the incident wave impedance. In order to realize this response, we consider structures with small, but finite, thickness, which allow engineering the required scattering matrix (4) with the desired nonlocal response. We consider transversely homogeneous thin multilayered planar slabs, in which each layer has a purely local response, but multiple reflections enable engineering the desired angular reflection and transmission spectra, with an approach analogous to the one recently employed to realize computational metastructures [32].

As an example, for bi-lateral excitation from TM-polarized sources with $\Delta\phi = \pi$, we chose to design a multilayered lossy structure having $R = -0.5$ and $T = 0.5$, consistent with Eqs. (4)-(5). This target design represents a strongly nonlinear inverse-design problem, with no analytical solution even for a few layers, which is best tackled relying on numerical optimization. Following this approach, as detailed in [17], we designed a thin symmetric stack of ten lossy layers that realizes the desired reflection and transmission coefficients, approximately constant as a function of angle. It should be stressed here that we considered no magnetic material, or other unconventional material properties in this design [17], making the fabrication of the structure feasible and practically appealing. Figs. 2a-b show the calculated reflection and transmission angular spectra, while Figs. 2c-d show the field distributions for different point-source illuminations (time-domain animations are provided in [17]). Despite being transversely homogeneous, when illuminated by localized point sources the designed structure operates as an omnidirectional CPA, absorbing the impinging radiation, only when illuminated from both sides with waves having the correct phase relation. Due to the translational symmetry, this response is independent of the source angular spectrum.

A full PT lens can be now obtained by pairing the designed passive CPA with its time-reversed counterpart, as sketched in Fig. 1b. This active element is obtained by conjugating the permittivity distribution obtained in the CPA design, such that $\varepsilon(\mathbf{r}) = \varepsilon^*(-\mathbf{r})$, where $\mathbf{r}$ is the position vector from the center of the PT lens. The scattering matrix of the complete system is indeed equal to Eq.

(1) [with $S_{22} = -2(e^{-2i\beta d} - 1)$], and the designed PT lens is therefore unidirectional invisible when illuminated from the CPA side, irrelevant of the impinging wavefront, and transmitted plane waves exhibit a *phase advance* $\Phi = -\beta d$ These properties are consistent with the findings in [16] for a pair of ultrathin PT-symmetric metasurfaces, but with the fundamental difference that here they are achieved *for any angle of incidence*, thanks to the implemented nonlocal response, and they are independent of the thickness $L$ of the active and passive elements. The numerically computed field distribution in Fig. 3 confirms that, under point-source illumination from the left side, the proposed PT lens indeed performs, in steady state, as an impedance-matched all-angle negative refraction slab, with negative phase and energy velocity, flowing from the active to the passive element (a time-domain animation is shown in [17]), consistent with the ray picture in Fig. 1b. We stress that this functionality works independent of the distance between passive and active elements, and simply assumes free-space in between the two structures. This represents the first example of all-angle negative refraction achieved in a *loss-immune, metamaterial-free, linear, and transversely invariant system*, with important implications for imaging, as we discuss in the following. In contrast with DNG lenses, in which material absorption is strongly detrimental for the performance of the device [17], here losses are at the very basis of the lensing mechanism, hence demonstrating the power of PT-symmetry concepts to go beyond the limitations of passive metamaterial structures.

## 4. IMAGING PROPERTIES OF PT-SYMMETRIC LENSES

As shown in Fig. 1b, and confirmed in Fig. 3, light rays emanating from a point source at distance $o$ from the PT lens are focused again at two distinct points, creating two *real images* of the source, the first in the central region of the lens, at a distance $o$ from the passive element, and the second outside the lens, at a distance $i = d - o$ from the active element. This imaging response is similar to the one of other systems based on negative refraction, but with significant differences. In fact, different from DNG and photonic crystal lenses, the first image can indeed be accessed for imaging purposes, as it is formed in free-space; besides, different from phase conjugating devices, the efficiency of the proposed PT lens may be significantly larger, since it is based on a linear scattering process. Another interesting feature of the proposed PT lens is that the relative phase of the central image can be controlled by the parameter $\Delta\phi$ of the CPA structure (it can be seen in Fig. 3 that the central image has a phase difference of $\Delta\phi = \pi$ with respect to the point source), an intriguing possibility for wave manipulation [33]. As expected, the images produced by the PT lens have diffraction-limited transverse resolution, because evanescent waves are not time-reversible, and therefore they do not contribute to the operation of a PT symmetric system. However, as seen in the inset of Fig. 3, the focus spot has transverse width $\approx 0.5\lambda_0$, where $\lambda_0$ is the free-space wavelength, which demonstrates that the designed PT lens is close to ideal from the point of view of geometrical optics. Moreover, as seen in Fig. 3, the PT lens offers good resolution in both transverse and longitudinal directions, different from the response of an ideal lossless DNG lens, in which subdiffractive transverse resolution inevitably comes at the cost of reduced longitudinal resolution, due to the strong resonant fields induced on the exit interface of the lens [10]-[11].

The imaging properties of the proposed PT lens are fundamentally different compared to any system of conventional lenses (i.e., based on conventional refraction). Remarkably, since the proposed setup is planar and transversely invariant, an ensemble of sources can be moved laterally in the object space without any distortion/aberration. Instead, while conventional lenses can be made planar, e.g., graded-refractive-index lenses, or, more recently, metasurface lenses [34]-[35], they can never be transversely homogenous, because they need to compensate, point by point, the phase difference of the light rays emerging from the source (conversely, PT lenses, or DNG lenses, automatically "rewind" the phase of the rays due to a negative phase velocity). In particular, conventional lenses have a definite optical axis, corresponding to the axis of rotational symmetry, and a definite focal distance $f$ that depends on the lens geometry. Conventional lens imaging is essentially based on the "thin-lens formula", $1/i + 1/o = 1/f$ [1], relating the position $i$ of the image, to the position $o$ of the source, through the focal distance. Instead, due to transverse invariance, PT lenses do not have an optical axis, are inherently *afocal*, and follow a different imaging rule, $i = d - o$, which implies that the position of the external image can be controlled by varying the distance $d$ between active and passive elements of the lens. In order to visually appreciate these different imaging properties, in Fig. 4 we compare the response of the designed PT lens (Fig. 4a) with that of a conventional spherical lens (Fig. 4b) under illumination from two point sources. Notably, since the two sources are located on different transverse planes, the spherical lens inevitably distorts the relative distance between their images, according to the thin-lens formula; conversely, due to its afocal nature, the PT lens always preserves relative distances and angles between points (as indicated by the blue arrows in Fig. 4), which is indeed one of the main characteristics of planar absolute optical instruments. Moreover, as clearly seen in Fig. 4, the image points produced by the PT lens are significantly sharper and more intense, whereas some degree of distortion/aberration (as well as reflection) is always expected for conventional lenses [1]-[3]. More generally, the intriguing properties of PT lenses may lead to the realization of aberration-free imaging devices, which may produce perfectly focused 3D images of a large region of space, of large interest for several practical applications.

## 5. SUMMARY

In conclusion, in this paper we have demonstrated that suitably designed PT-symmetric systems, combined with a tailored nonlocal response, allow realizing a loss-immune, metamaterial-free, linear, transversely invariant, aberration-free lens, which implements the imaging properties of planar absolute optical instruments without some of the limitations of existing metamaterial lenses. In particular, the negative phase and power flow distributions support an imaging process analogous to the one of an ideal plane mirror, as seen in Fig. 1, but with a real image space. This represents an interesting parallel between PT-lens imaging, which is based on space-time *mirror* symmetry, and conventional mirror imaging. It should also be noted that, since the proposed PT lens is afocal, it cannot perform image magnification, which may be a useful feature in many applications. To circumvent this limitation, we are currently exploring PT-inspired cylindrical lenses that may allow realizing a general platform for imaging and wave manipulation. While a proof-of-concept realization of these concepts may be readily envisioned at microwave frequencies using suitably designed stacks of passive and active metasurfaces, at optical frequencies the use of parametric gain may be explored to realize low-noise gain elements. We have confirmed that the proposed active system can be made stable over a finite bandwidth, consistent with the experimental results in [25]. We believe that the novel concept of aberration-free PT imaging, put forward in this paper, may open uncharted directions in the field of optical imaging. This work was supported by the Air Force Office of Scientific Research. C.A.V. was partially supported by the Academy of Finland under postdoctoral project funding 13260996.

**References**

[1] M. Born and E. Wolf, *Principles of Optics: Electromagnetic Theory of Propagation, Interference and Diffraction of Light* (Cambridge University Press, 1999).

[2] J. C. Maxwell, "On the general laws of optical instruments," Q. J. Pure Appl. Math. II, 271-285 (1858).

[3] A. Walther, "Irreducible aberrations of a lens used for a range of magnifications," J. Opt. Soc. Am. A **6**, 415 (1989).

[4] T. Smith, "On perfect optical instruments," Proc. Phys. Soc. **60**, 293 (1948).

[5] Interestingly, while a plane mirror was the only known example of planar absolute optical instrument before the advent of metamaterials, different non-planar spherically symmetric absolute instruments are well-known in classical optics, most notably the Maxwell's fish-eye lens, variants of the Luneberg lens, etc. [1],[6].

[6] T. Tyc, L. Herzánová, M. Šarbort, and K. Bering, "Absolute instruments and perfect imaging in geometrical optics," New J. Phys. **13**, 115004 (2011).

[7] V. G. Veselago, "The Electrodynamics of Substances with Simultaneously Negative Values of ε and μ" Sov. Phys. Uspekhi **10**, 509 (1968).

[8] J. B. Pendry, "Negative Refraction Makes a Perfect Lens," Phys. Rev. Lett. **85**, 3966 (2000).

[9] V. A. Podolskiy and E. E. Narimanov, "Near-sighted superlens," Opt. Lett. **30**, 75 (2005).

[10] X.-X. Liu and A. Alu, "Limitations and potentials of metamaterial lenses," J. Nanophotonics **5**, 053509 (2011).

[11] A. Grbic and G. Eleftheriades, "Overcoming the Diffraction Limit with a Planar Left-Handed Transmission-Line Lens," Phys. Rev. Lett. **92**, 117403 (2004).

[12] S. Maslovski and S. Tretyakov, "Phase Conjugation and Perfect Lensing," J. Appl. Phys. **94**, 4241 (2003).

[13] J. B. Pendry, "Time Reversal and Negative Refraction," Science **322**, 71 (2008).

[14] M. Notomi, "Theory of light propagation in strongly modulated photonic crystals: Refractionlike behavior in the vicinity of the photonic band gap," Phys. Rev. B **62**, 10696 (2000).

[15] C. Luo, S. Johnson, J. Joannopoulos, and J. Pendry, "All-angle negative refraction without negative effective index," Phys. Rev. B **65**, 201104 (2002).

[16] R. Fleury, D. L. Sounas, and A. Alù, "Negative Refraction and Planar Focusing Based on Parity-Time Symmetric Metasurfaces," Phys. Rev. Lett. **113**, 023903 (2014).

[17] See Supplementary Material for the derivation and discussion of Eqs. (2)-(5), further details on the design of omnidirectional CPA structures, as well as a comparison between the designed PT lens and a DNG lens with realistic losses. Supplementary Material also includes time-domain animations of the field distributions in Figs. 2-3.

[18] C. M. Bender and S. Boettcher, "Real Spectra in Non-Hermitian Hamiltonians Having PT Symmetry," Phys. Rev. Lett. **80**, 5243 (1998).

[19] K. G. Makris, R. El-Ganainy, D. N. Christodoulides, and Z. H. Musslimani, "Beam Dynamics in PT Symmetric Optical Lattices," Phys. Rev. Lett. **100**, 103904, (2008).

[20] C. E. Rüter, K. G. Makris, R. El-Ganainy, D. N. Christodoulides, M. Segev, and D. Kip, "Observation of parity–time symmetry in optics," Nat. Phys. **6**, 192–195 (2010).

[21] S. Longhi, "PT-symmetric laser absorber," Phys. Rev. A **82**, 031801 (2010).

[22] Z. Lin, H. Ramezani, T. Eichelkraut, T. Kottos, H. Cao, and D. N. Christodoulides, "Unidirectional Invisibility Induced by PT-Symmetric Periodic Structures," Phys. Rev. Lett. **106**, 213901 (2011).

[23] H. Hodaei, M.-A. Miri, M. Heinrich, D. N. Christodoulides, and M. Khajavikhan, "Parity-time-symmetric microring lasers," Science **346** 975–8 (2014).

[24] X. Zhu, H. Ramezani, C. Shi, J. Zhu, and X. Zhang, "PT-Symmetric Acoustics," Phys. Rev. X **4**, 031042 (2014).

[25] R. Fleury, D. L. Sounas, and A. Alù, "An invisible acoustic sensor based on parity-time symmetry.," Nat. Commun. **6**, 5905 (2015).

[26] D. L. Sounas, R. Fleury, and A. Alù, "Unidirectional Cloaking Based on Metasurfaces with Balanced Loss and Gain," Phys. Rev. Applied **4**, 014005 (2015).

[27] Y. D. Chong, L. Ge, H. Cao, and A. D. Stone, "Coherent Perfect Absorbers: Time-Reversed Lasers," Phys. Rev. Lett. **105**, 053901 (2010).

[28] W. Wan, Y. Chong, L. Ge, H. Noh, A. D. Stone, and H. Cao, "Time-reversed lasing and interferometric control of absorption," Science **331**, 889–92 (2011).

[29] C. M. Watts, X. Liu, and W. J. Padilla, "Metamaterial electromagnetic wave absorbers," Adv. Mater. **24**, OP98–120 (2012).

[30] C. Argyropoulos, "Electromagnetic Absorbers Based on Metamaterial and Plasmonic Devices", Forum for Electromagnetic Research Methods and Application Technologies (FERMAT) **2**, (2014).

[31] C. A. Valagiannopoulos, A. Tukiainen, T. Aho, T. Niemi, M. Guina, S. A. Tretyakov, and C. R. Simovski, "Perfect magnetic mirror and simple perfect absorber in the visible spectrum," Phys. Rev. B **91**, 115305 (2015).

[32] A. Silva, F. Monticone, G. Castaldi, V. Galdi, A. Alù, and N. Engheta, "Performing mathematical operations with metamaterials," Science **343**, 160–3 (2014).

[33] It should be noted that only a phase difference $\Delta\phi = 0, \pi, 2\pi, \ldots$ guarantees ideal imaging in the sense of Eq. (1), consistent with the fact that a symmetric structure can realize CPA operation only for symmetric or anti-symmetric illumination, as shown in [17] and [27]. The generalization of these results to non-symmetric CPAs and arbitrary phase delay $\Delta\phi$, which may realize other interesting wave-manipulation functionalities (different from ideal imaging), will be the subject of a future work.

[34] F. Monticone, N. M. Estakhri, and A. Alù, "Full Control of Nanoscale Optical Transmission with a Composite Metascreen," Phys. Rev. Lett. **110**, 203903 (2013).

[35] N. Yu and F. Capasso, "Flat optics with designer metasurfaces," Nat. Mater. **13**, 139–50 (2014).

**Figures**

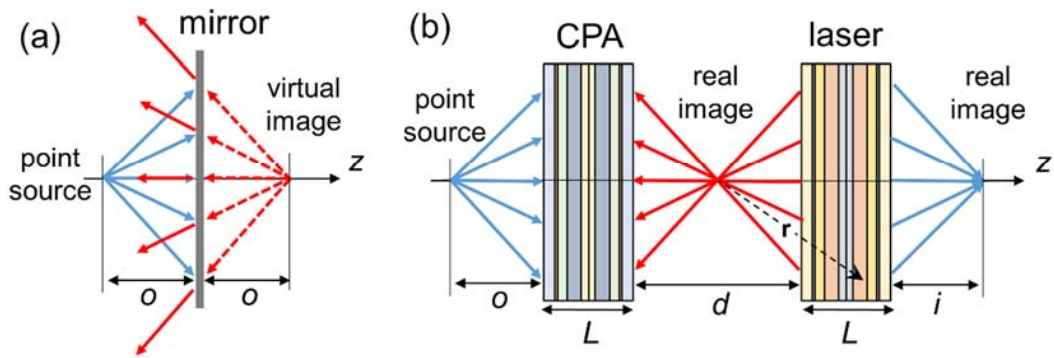

Figure 1 – Comparison of aberration-free imaging, under point-source illumination, by (a) a plane mirror and (b) the proposed nonlocal PT-symmetric lens, composed of an omnidirectional coherent perfect absorber (CPA) paired with its time-reversed counterpart. In both cases, blue and red arrows denote waves propagating forward and backward, respectively, with respect to the positive z axis. Dashed arrows in (a) indicate rays originating from the virtual image.

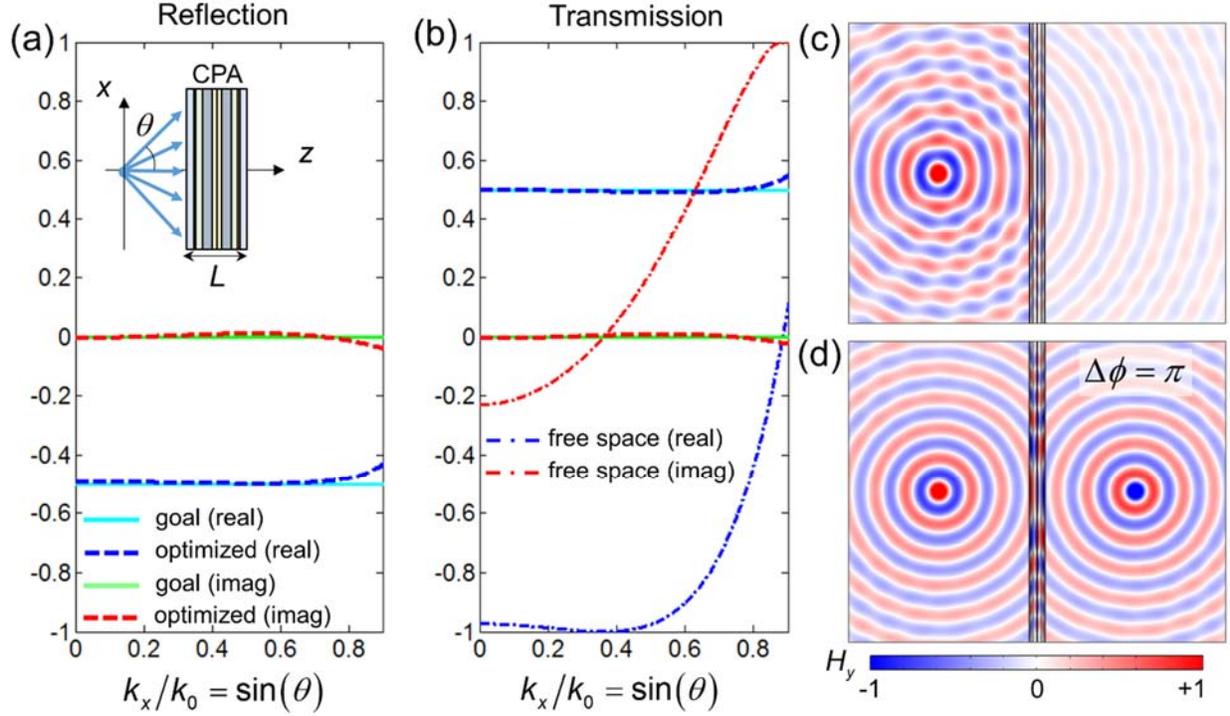

Figure 2 – Omnidirectional CPA based on a multilayered planar slab [inset of panel (a)]. (a) Reflection and (b) transmission coefficients, as a function of the normalized transverse wavenumber $k_x/k_0$, for the optimized structure (dashed lines; red, imaginary part; blue, real part), and goal functions $R = -0.5$ and $T = 0.5$ (solid lines; green, imaginary part; cyan, real part). The total length of the optimized CPA is only $L = 0.537\lambda_0$, where $\lambda_0$ is the free-space wavelength (details in [17]). For comparison, in panel (b) we also show the transmission coefficient (dot-dashed lines) of an equivalent free-space slab having same length $L$, i.e., $T_{fs}(k_x) = \exp\left(i\sqrt{k_0^2 - k_x^2}L\right)$, which highlights the challenge of realizing angle-independent scattering parameters. (c-d) Magnetic field distribution (time-snapshot) of the designed omnidirectional CPA under illumination from (a) single TM-polarized point source and (b) two symmetrically located point sources with phase difference $\Delta\phi = \pi$ (animations in [17]).

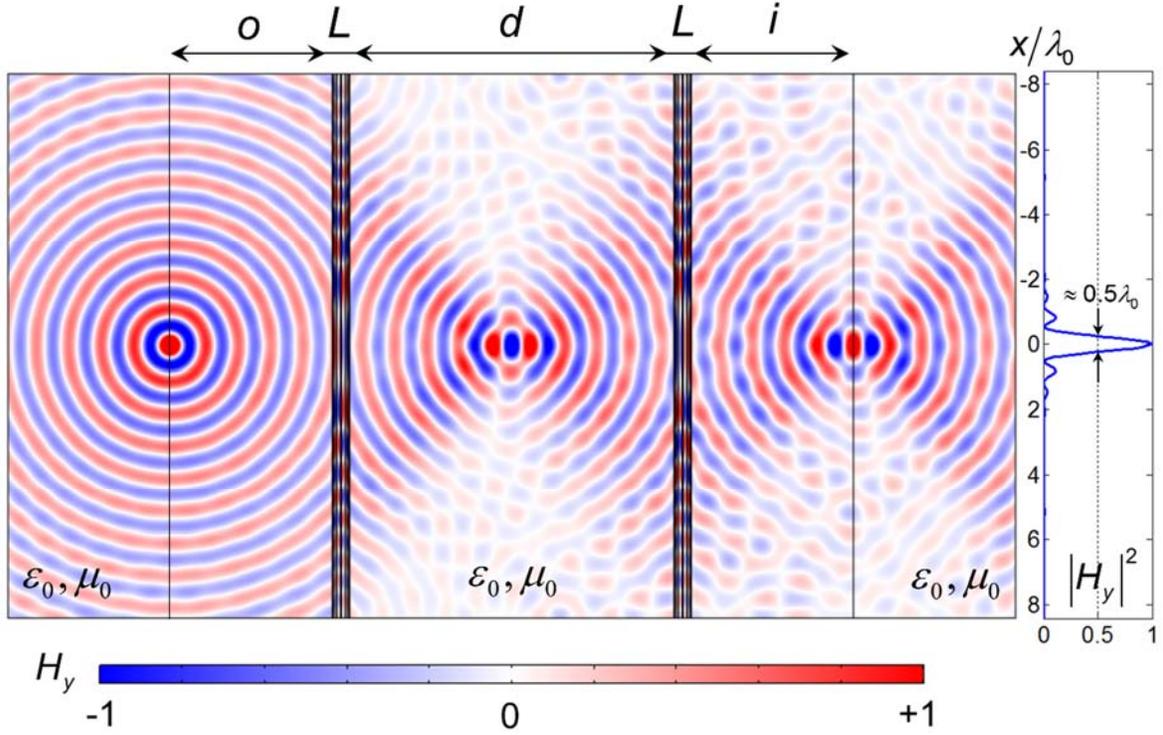

Figure 3 – Magnetic field distribution (time-snapshot) of the proposed PT lens, under TM-polarized point source illumination on the left side (animation in [17]). The lens is homogenous along the transverse $x$ direction. The inset on the right shows the field intensity on the focal plane (at distance $i$ from the active element), demonstrating transverse resolution very close to the diffraction limit.

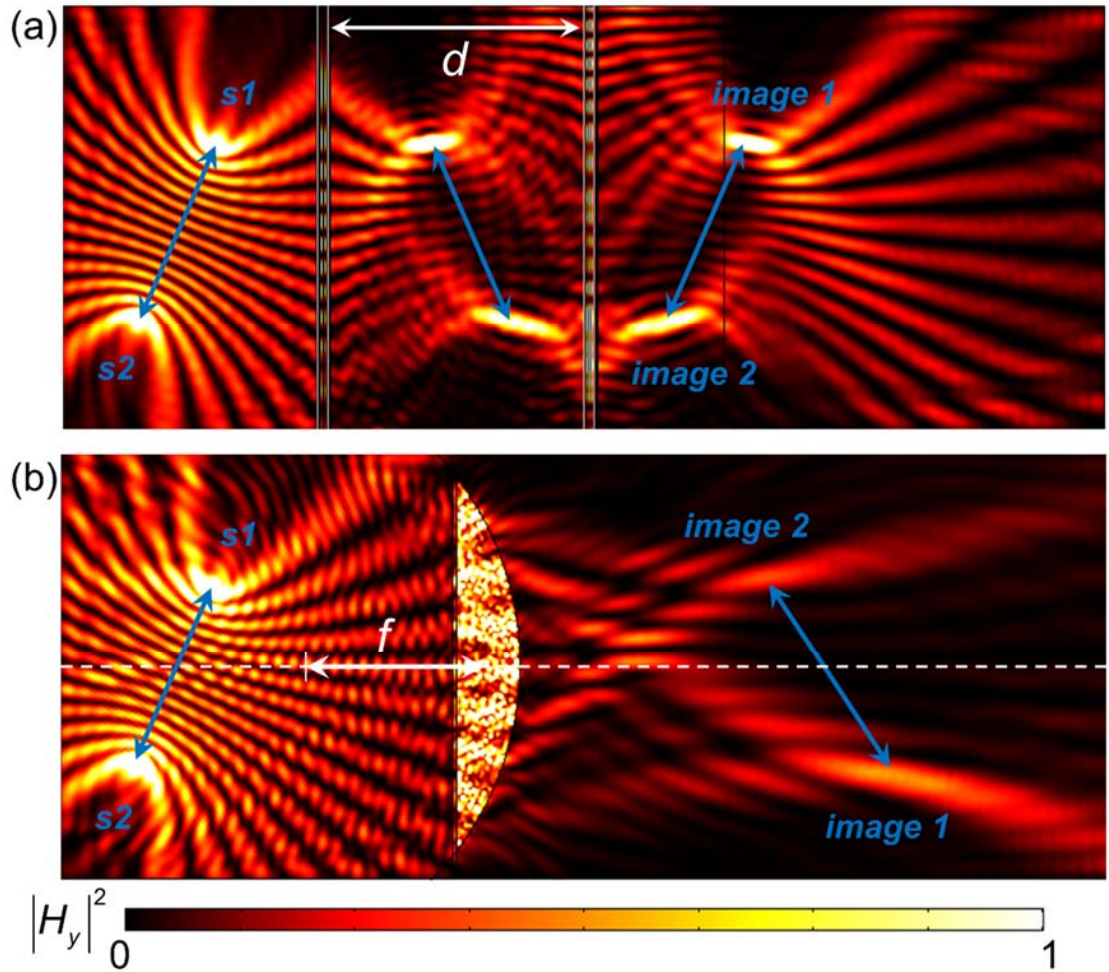

Figure 4 – Comparison between the field intensity distribution of (a) the proposed PT lens, with length *d*, and (b) a conventional spherical dielectric lens (refractive index $n = 2.5$) with focal length *f* , under illumination from two point sources *s*1 and *s*2 located on different transverse planes. The blue arrows indicate the relative position of sources and images, highlighting the different imaging properties of the two lenses.

Supplementary material for the paper

"Aberration-Free Imaging Based on Parity-Time Symmetric Nonlocal Metasurfaces"

Francesco Monticone, Constantinos A. Valagiannopoulos and Andrea Alù

**1. Derivation of PT-Symmetry Conditions for Ideal Aberration-Free Imaging**

Eq. (1) of the main text represents the scattering matrix of a generic system that performs ideal aberration-free imaging. In order to design a structure that realizes this scattering response, it is useful to use a transfer-matrix formalism, in which the cascade of different blocks can be represented as the cascade (matrix product) of their transfer matrices. The transfer matrix corresponding to Eq. (1) can readily be obtained (see, e.g., Ref. [S1]):

$$T = \begin{pmatrix} \cos(\beta d) - \frac{1}{2} S_{22} e^{i\beta d} & +i Z_0 \sin(\beta d) + \frac{1}{2} Z_0 S_{22} e^{i\beta d} \\ +i \sin(\beta d)/Z_0 - \frac{1}{2Z_0} S_{22} e^{i\beta d} & \cos(\beta d) - \frac{1}{2} S_{22} e^{i\beta d} \end{pmatrix}, \quad (S1)$$

which has unitary determinant and, if $S_{22} = 0$, it becomes equal to the transfer matrix of a transmission-line of length $d$ and negative phase velocity $v_p = \omega/\beta = -c_0/\cos\theta$ (where $c_0$ is the speed of light in vacuum, and $\theta$ is the incidence angle), exactly the opposite compared to a free-space region of same length.

Here, we would like to implement Eq. (S1) as the cascade of three blocks, as shown in Fig. S1: two elements with transfer matrices $T_L$ and $T_R$, separated by an air region of length $d$ modeled as a transmission line with transfer matrix:

$$T_{line} = \begin{pmatrix} \cos(\beta d) & -iZ_0 \sin(\beta d) \\ -i\sin(\beta d)/Z_0 & \cos(\beta d) \end{pmatrix}. \tag{S2}$$

Our goal is therefore to determine $T_L$ and $T_R$ such that:

$$T = T_L T_{line} T_R. \tag{S3}$$

To solve this problem, we perform an *eigendecomposition* of the square matrices $T$ and $T_{line}$, which allows writing them in terms of their eigenvalues and eigenvectors as

$$\begin{aligned} T &= V \Lambda V^{-1} \\ T_{line} &= V_{line} \Lambda_{line} V_{line}^{-1} \end{aligned}, \tag{S4}$$

where the columns of $V$ and $V_{line}$ are the eigenvectors of the transfer matrices, whereas $\Lambda$ and $\Lambda_{line}$ are diagonal matrices whose elements are the corresponding eigenvalues. With this decomposition, we can write Eq. (S3) as

$$V \Lambda V^{-1} = T_L \left( V_{line} \Lambda_{line} V_{line}^{-1} \right) T_R. \tag{S5}$$

Interestingly, the matrices $T$ and $T_{line}$ have the same eigenvalues, therefore

$$\Lambda = \Lambda_{line} = \begin{pmatrix} e^{i\beta d} & 0 \\ 0 & e^{-i\beta d} \end{pmatrix}. \tag{S6}$$

[Note that $\Lambda = \Lambda_{line}$ only if we assume that the corresponding structures have identical length $d$; however, if the structures have different lengths, say $d$ and $d_l$, we can write

$$\begin{aligned} \Lambda_{line} &= diag\left(e^{\pm i\beta d_l}\right) = diag\left(e^{\pm i\beta(d_l-d)/2}\right) \cdot diag\left(e^{\pm i\beta d}\right) \cdot diag\left(e^{\pm i\beta(d_l-d)/2}\right) = \\ &= diag\left(e^{\pm i\beta(d_l-d)/2}\right) \cdot \Lambda \cdot diag\left(e^{\pm i\beta(d_l-d)/2}\right) \end{aligned} \tag{S7}$$

and the following derivation is only slightly modified].

By applying Eq. (S6) to (S5) it follows that

$$\begin{cases} V = T_L V_{line} \\ V^{-1} = V_{line}^{-1} T_R \end{cases}, \quad (S8)$$

and, since $V$ and $V_{line}$ are known, we can find the transfer matrices of the left and right elements by simply inverting Eq. (S8):

$$\begin{cases} T_L = V V_{line}^{-1} \\ T_R = V_{line} V^{-1} \end{cases}. \quad (S9)$$

Note also that, since the inverse of a matrix product is anti-commutative, the left and right transfer matrices are one the inverse of the other, i.e.,

$$T_R = T_L^{-1}. \quad (S10)$$

Finally, we build the matrices $V$ and $V_{line}$ from the eigenvectors of the corresponding transfer matrices:

$$V_{line} = \begin{pmatrix} a_l & b_l \\ \dfrac{a_l}{Z_0} & -\dfrac{b_l}{Z_0} \end{pmatrix}, \quad V = \begin{pmatrix} a & b \\ a \dfrac{S_{22} \cos(\beta d) + i(-2 + S_{22})\sin(\beta d)}{S_{22} Z_0 \cos(\beta d) + i(2 + S_{22}) Z_0 \sin(\beta d)} & \dfrac{b}{Z_0} \end{pmatrix}, \quad (S11)$$

where $a, b, a_l, b_l \in \mathbb{C}$ are arbitrary non-zero complex constants (for any value of these constants, the columns of the matrices above represent eigenvectors of the corresponding transfer matrices). Eq. (S11) can be substituted into (S9) to find the transfer matrices $T_L = T_R^{-1}$, whose general expression is not written here for the sake of brevity. The corresponding scattering matrix of the left element can be written as

$$S_L = \begin{pmatrix} \dfrac{2i}{S_{22}\left[i+\cot(\beta d)\right]} & -\dfrac{2bi}{b_l S_{22}\left[i+\cot(\beta d)\right]} \\ \dfrac{a_l}{a}\left(1+\dfrac{2i}{S_{22}\left[i+\cot(\beta d)\right]}\right) & -\dfrac{a_l b\left[i(2+S_{22})+S_{22}\cot(\beta d)\right]}{ab_l S_{22}\left[i+\cot(\beta d)\right]} \end{pmatrix}. \tag{S12}$$

which represents the most general form of $S_L$ to realize a system, as in Fig. S1b, with overall $S$ matrix given by Eq. (1) of the main text, namely, a system that performs ideal imaging. Eq. (S12) represents a structure that is generally non-reciprocal, non-symmetric and that can absorb or emit energy (the scattering matrix is non-unitary). However, the structure can easily be made *reciprocal* by imposing:

$$S_{22} = -2i\dfrac{ab+a_l b_l}{a_l b_l \left[i+\cot(\beta d)\right]} . \tag{S13}$$

If we further require the left element to be *symmetric*, namely, with identical reflection coefficient from both sides, we obtain the following condition:

$$b = \pm b_l, \tag{S14}$$

which should hold for any $a, a_l, b, b_l \in \mathbb{C}$. If, for example, we choose the plus sign in (S14), and substitute it in Eq. (S12) together with the reciprocity condition (S13), we obtain

$$S_L = \begin{pmatrix} -c & c \\ c & -c \end{pmatrix}, \tag{S15}$$

where we introduced a new complex constant $c = a_l/(a+a_l)$. With these assumptions, the corresponding transfer matrices become

$$T_L = T_R^{-1} = \begin{pmatrix} \dfrac{1}{2c} & \dfrac{Z_0}{2}\left(\dfrac{1-2c}{c}\right) \\ \dfrac{1}{2Z_0}\left(\dfrac{1+2c}{c}\right) & \dfrac{1}{2c} \end{pmatrix}, \qquad (S16)$$

which can be implemented with a suitably designed symmetric T or Π circuit network [S1]. In particular, if we choose $c = 1/2$, the reciprocity condition for $S_L$, given by Eq. (S13), simplifies to $S_{22} = -2\left(1 - e^{-2i\beta d}\right)$, and the transfer matrices become

$$T_L = \begin{pmatrix} 1 & 0 \\ \dfrac{1}{Z_0/2} & 1 \end{pmatrix}, \quad T_R = \begin{pmatrix} 1 & 0 \\ -\dfrac{1}{Z_0/2} & 1 \end{pmatrix}, \qquad (S17)$$

which represent a parity-time symmetric pair of *shunt-impedance sheets*, or metasurfaces, with impedance value $R = \pm Z_0/2$. Eq. (S17) correspond to Eq. (3) of the main text.

Instead, if we choose the minus sign in the symmetry condition (S14), the scattering matrix of the left element becomes

$$S_L = \begin{pmatrix} c & c \\ c & c \end{pmatrix}, \qquad (S18)$$

where now $c = a_t/(a - a_t)$, and the corresponding transfer matrices are found to be

$$T_L = T_R^{-1} = \begin{pmatrix} \dfrac{1}{2c} & \dfrac{Z_0}{2}\left(\dfrac{1+2c}{c}\right) \\ \dfrac{1}{2Z_0}\left(\dfrac{1-2c}{c}\right) & \dfrac{1}{2c} \end{pmatrix}. \qquad (S19)$$

Eqs. (S16) and (S19) together correspond to Eq. (2) of the main text, and Eqs. (S15) and (S18) to Eq. (4). Again, if we choose $c = 1/2$ in (S19), the reciprocity condition for $S_L$, given by Eq. (S13), simplifies to $S_{22} = +2(1 - e^{-2i\beta d})$, and the transfer matrices become

$$T_L = \begin{pmatrix} 1 & 2Z_0 \\ 0 & 1 \end{pmatrix}, \quad T_R = \begin{pmatrix} 1 & -2Z_0 \\ 0 & 1 \end{pmatrix}, \quad (S20)$$

which are exactly the dual of Eqs. (S17), and represent a pair of *series-impedance sheets* of value $R = \pm 2Z_0$.

As mentioned in the main text, and further discussed in the following section, both Eqs. (S15) and (S18) represent the scattering matrix of a symmetric and reciprocal coherent perfect absorber (CPA) [however, all this discussion can readily be extended, in the most general scenario, to CPAs that are both non-symmetric and non-reciprocal]. In particular, the fact that symmetry determines two possible forms for the scattering matrix of the left element, i.e., Eqs. (S15) or (S18), is consistent with the fact that, in the case of a symmetric structure, ideal CPA operation can be achieved *only* for symmetric or anti-symmetric excitation, corresponding to either Eq. (S15) or Eq. (S18) according to whether the excitation is TE- or TM-polarized, as we will see in the next section.

Eqs. (S15) and (S18) give the scattering matrix of the left element of the system in Fig. S1b, assuming this element is reciprocal and symmetric. Interestingly, it can easily be verified that the right element of the system is instead characterized by a non-finite $S$ matrix, consistent with the fact that the right element is an active, coherently emitting structure (i.e., the time-reversed counterpart of a CPA). Nevertheless, the scattering matrix of the entire system is finite, and is given by

$$S = \begin{pmatrix} 0 & e^{-i\beta d} \\ e^{-i\beta d} & \mp \dfrac{1-e^{-2i\beta d}}{c} \end{pmatrix}, \tag{S21}$$

in which the minus (plus) sign corresponds to the plus (minus) sign in Eq. (S14), and c is again a generic complex constant. Eq. (S21) is indeed of the form of Eq. (1) of the main text.

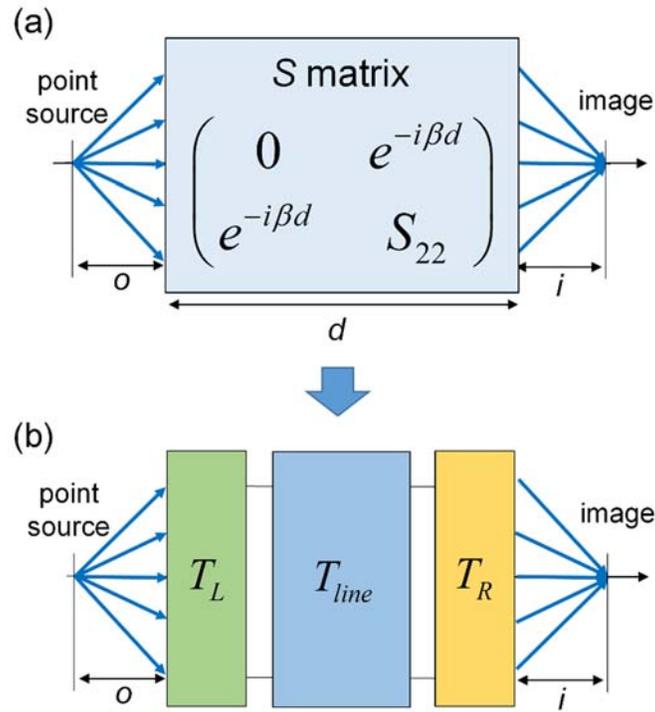

Figure S1 – (a) Generic structure that realizes ideal aberration-free imaging, according to the scattering matrix $S$ in Eq. (1) of the main text, and (b) desired implementation composed of the cascade of three blocks, characterized by transfer matrices $T_L$, $T_{line}$ and $T_R$.

## 2. Design of Omnidirectional Coherent Perfect Absorbers

The scattering-matrix formalism allows obtaining the vector of outgoing waves $b$ at the different "ports" of a structure from the vector of input waves $a$ through the scattering matrix of the structure. For a generic slab, illuminated from both sides with waves having equal amplitude $|a| \neq 0$ and relative phase different $\Delta\phi$, as illustrated in Fig. S1, the scattering-parameter system of equations becomes:

$$\begin{bmatrix} b_1 \\ b_2 \end{bmatrix} = \begin{bmatrix} S_{11} & S_{12} \\ S_{21} & S_{22} \end{bmatrix} \begin{bmatrix} a \\ a e^{i\Delta\phi} \end{bmatrix}. \tag{S22}$$

CPA operation implies that the outgoing waves need to completely vanish, i.e., $b_1 = b_2 = 0$. The trivial solution with all scattering parameters equal to zero (vanishing reflection and transmission coefficients), however, needs to be excluded, since it corresponds to a conventional perfect absorber, in which perfect absorption under unilateral excitation is achieved independently of what happens at the other port (which can indeed be substituted by a short circuit, as in grounded absorbing slabs). Instead, in a CPA, perfect absorption is obtained as the result of interference between reflected and transmitted waves under bi-lateral excitation. Therefore, the conditions for CPA operation read

$$\begin{aligned} S_{11} + S_{12} e^{i\Delta\phi} &= 0 \\ S_{21} + S_{22} e^{i\Delta\phi} &= 0 \end{aligned}. \tag{S23}$$

If we assume that the structure is reciprocal, i.e., $S_{12} = S_{21} = T$, and symmetric, i.e., $S_{11} = S_{22} = R$, as in the case considered in the main text and in the previous section, Eqs. (S23) become

$$\begin{aligned} R + T e^{i\Delta\phi} &= 0 \\ T + R e^{i\Delta\phi} &= 0 \end{aligned}, \tag{S24}$$

which can be simultaneously true only if $\Delta\phi = 0, \pi, 2\pi, ...$ . In other words, because of the mirror symmetry of the structure, CPA operation can be achieved only for symmetric or anti-symmetric excitation, as mentioned in the previous section and discussed in Ref. [27]. Instead, non-symmetric structures can realize, in principle, coherent absorption for arbitrary phase difference $\Delta\phi$, with interesting implications for imaging and wave manipulations, as we will discuss in a future paper. It is also important to note that the quantities $a_i$ and $b_i$ are typically defined as voltage waves in an equivalent transmission line, which models the longitudinal wave propagation in the considered system (see, e.g., Ref. [S1]). In particular, the voltage corresponds to the transverse component of the electric field of the propagating wave. Therefore, if the sources illuminating the structure in Fig. S1 are *TE-polarized* (transverse out-of-plane electric field), the condition for CPA operation directly follows from the above equations:

$$R + Te^{i\Delta\phi} = 0, \quad \Delta\phi = 0, \pi. \tag{S25}$$

Instead, if the sources are defined in terms of a transverse magnetic field, which corresponds to the current in the equivalent transmission-line model, Eq. (S25) needs to be slightly modified to account for the minus sign in the definition of current waves propagating in the $-z$ direction [S1]:

$$R - Te^{i\Delta\phi} = 0, \quad \Delta\phi = 0, \pi. \tag{S26}$$

which corresponds to Eq. (4) of the main text for *TM-polarized* illumination. The sign difference in (S25) and (S26) is consistent with the fact that, when a CPA structure is illuminated by anti-symmetric TM-polarized sources (i.e., $\Delta\phi = \pi$), as in Fig. 2d of the main text, the corresponding transverse electric fields ($E_x$ component), impinging from the two sides, are actually in-phase.

In light of this discussion, it is therefore clear that the left element of the structure in Fig. S1, whose scattering matrix is given by Eq. (S15) or (S18), is indeed a CPA structure with $|R|=|T|=c$. In particular, Eq. (S15) corresponds to the case of symmetric TE excitation [condition (S25) with $\Delta\phi=0$, as considered in Ref. [16]] or the case of anti-symmetric TM excitation [condition (S26) with $\Delta\phi=\pi$, as considered in the main text]. Conversely, Eq. (S18) corresponds to the case of anti-symmetric TE excitation [condition (S25) with $\Delta\phi=\pi$] or the case of symmetric TM excitation [condition (S26) with $\Delta\phi=0$].

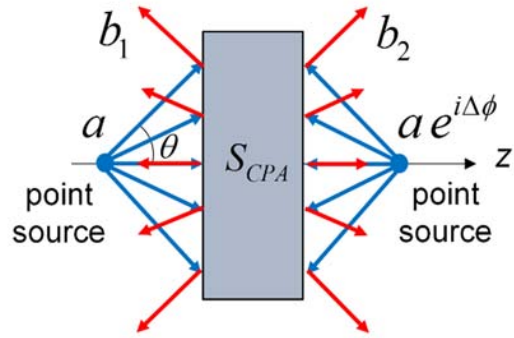

Figure S2 – Sketch of a generic slab described by a two-port scattering matrix $S_{CPA}$, with input incoming waves $a$ and $ae^{i\Delta\phi}$ (blue arrows), and output outgoing waves $b_1$ and $b_2$ (red arrows), consistent with Eq. (S22). For CPA operation, the outgoing waves need to completely vanish, as a result of interference and dissipation in the structure.

From the physical standpoint, Eq. (S25), or (S26), indicates that the reflected wave from one side of the structure destructively interferes with the wave transmitted from the other side, such that the field is confined in an interference pattern within the CPA structure and gets completely absorbed, as discussed in Ref. [25]. In general, however, this effect occurs only at a specific incidence angle, or within a narrow angular range, as the reflection and transmission coefficients may strongly vary as a function of angle. Therefore, to achieve omnidirectional coherent absorption, for any bilateral source distribution, we further require $R$ and $T$ to be angle-independent. In the main text, we have considered the case of TM-polarized bilateral illumination with $\Delta\phi = \pi$, which, according to the above discussion, requires a CPA structure with scattering matrix of the form of Eq. (S15). In particular, we have chosen to implement a scattering matrix as in Eq. (S15) with $c = 1/2$:

$$S_{CPA} = \frac{1}{2}\begin{bmatrix} -1 & 1 \\ 1 & -1 \end{bmatrix} \quad \forall \theta. \tag{S27}$$

It should be noted that the specific values of $R$ and $T$ (i.e., the value of $c$) can be chosen arbitrarily, as long as Eq. (S25), or (S26), is fulfilled (even if they lead to a non-passive structure). The only forbidden values for $R$ and $T$ are the ones that lead to an ideally lossless structure (i.e., any complex $R$ and $T$ for which $|R|^2 + |T|^2 = 1$), since a finite amount of absorption is always necessary for CPA operation.

The design a multilayered slab that implements the required scattering parameters (S27) is an example of strongly nonlinear inverse problem. To tackle this design problem, we developed a fast synthesis procedure based on nonlinear optimization techniques, analogous to the one used in Ref. [32] (see supplementary material therein). In particular, the optimization algorithm is based on the Nelder-Mead simplex optimization method (available in Matlab as the function *fminsearch*), and the starting point was moved across a large range of parameters, searching for a sufficiently good solution in different regions of the multi-dimensional parameter space (consider that here we are not interested in finding the global minimum).

We considered a multilayered slab composed of ten symmetric layers (thereby, only half of the parameters were subject to optimization), with no magnetic properties, i.e., with free-space magnetic permeability, and we aimed at minimizing the difference between the angular transmission and reflection spectra of the multilayered slab and the target scattering parameters defined by Eq. (S27). In particular, the error function was defined as the mean square error, for both real and imaginary parts of the parameters, at discrete equispaced points within the range $0 < k_x < k_0$ (where $k_x = k_0 \sin\theta$ is the transverse wavenumber), namely, within the propagative portion of the angular spectrum. Moreover, in order to obtain realistic values for the involved parameters, we constrained the real part of the relative permittivity within the range $-50 < \text{Re}[\varepsilon] < 50$ and the layer thickness $d > \lambda_0/100$ (i.e., layers not too thin). In addition, we aimed at minimizing the imaginary part of the permittivity of the different layers, such that the operation of the PT lens is supported by the minimum amount of loss and gain, hence relaxing the issues of instability and large spontaneous emission that may arise for large level of gain, particularly at optical frequencies.

Table S1 reports the relative permittivity and layer thickness of the first 5 layers of the optimized CPA structure (the remaining 5 layers are symmetric). Notably, all the obtained parameter values are realistic, hence making the fabrication of the structure feasible. The scattering parameters of the optimized structure, reported in Figs. 2a-b of the main text, are approximately constant with the angle of incidence and closely match the required response. This demonstrates the possibility of realizing an omnidirectional CPA (and therefore a planar PT lens) without the need of unconventional material properties.

|   | layer 1 | layer 2 | layer 3 | layer 4 | layer 5 |
|---|---|---|---|---|---|
| $\varepsilon$ | $22.50 + 0.80i$ | $3.50$ | $6.90$ | $-2.55 + 0.13i$ | $15.60 + 0.60i$ |
| $d$ | $\lambda_0/34.0$ | $\lambda_0/17.0$ | $\lambda_0/17.5$ | $\lambda_0/9.1$ | $\lambda_0/73.8$ |

Table S1 – Relative electric permittivity and layer thickness of the first 5 layers of the optimized omnidirectional CPA. We assumed a $e^{-i\omega t}$ time-harmonic convention; therefore, positive imaginary part corresponds to losses.

## 2. Lossy DNG Lens vs. PT Lens

One of the main advantages of the proposed PT lens is its loss-immune response. More specifically, material losses are a fundamental ingredient of the PT lensing mechanism. Conversely, in typical metamaterial flat lenses, such as DNG lenses, material absorption is strongly detrimental for the performance of the device; therefore, in conventional metamaterial implementations, the goal is always to reduce losses as much as possible. Unfortunately, since metamaterials are typically based on the resonant response of their subwavelength inclusions, material absorption is unavoidable, because losses are fundamentally associated to frequency dispersion through Kramers-Kronig relations [S2]. In order to visually appreciate the effect of absorption, in Fig. S3 we compare the field intensity distribution of the PT lens presented in the main text with that of a lossy DNG lens. In particular, we chose the relative permittivity and permeability of the DNG material as $\varepsilon_{DNG} = -1+0.02i$ and $\mu_{DNG} = -1+0.02i$, in which the imaginary part has been chosen to be similar to that of the permittivity of silver, when its real part is -1. It should be noted that this is an "optimistic" choice for the metamaterial losses, since silver has arguably the lowest level of losses among materials with small and negative permittivity at optical frequencies [S3]. Nevertheless, it is clear from Fig. S3b that material absorption almost completely suppresses the image produced by the DNG lens, as the fields strongly decay inside the material (besides, subwavelength resolution is not achievable in this case, since also the evanescent spectrum decays inside the DNG slab). Conversely, in the proposed PT lens, thanks to the interplay of loss and gain, the intensity of the image is very high, comparable to the intensity of the source, as seen in Fig. S3a.

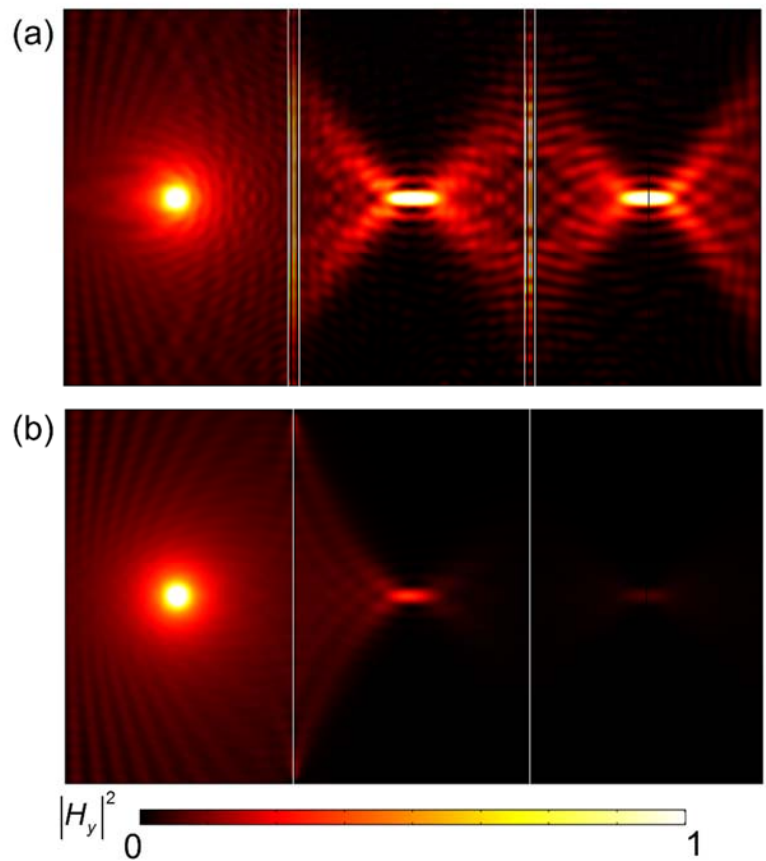

Figure S3 – Field intensity distribution of (a) the proposed PT lens, as in Figs. 3-4 of the main text, and (b) a lossy DNG lens (located between the two vertical white lines).

### 3. Animations

Supplementary material also includes the following attached animations:

- **Animation_S1_cpa.gif**: time-domain animation of the magnetic field distribution for the designed CPA under unilateral point-source excitation, corresponding to Fig. 2c.

- **Animation_S2_cpa.gif**: time-domain animation of the magnetic field distribution for the designed CPA under bi-lateral anti-symmetric excitation, corresponding to Fig. 2d.
- **Animation_S3_cpa.gif**: time-domain animation of the magnetic field distribution for the designed CPA under bi-lateral symmetric excitation.
- **Animation_S4_PT_lens.gif**: time-domain animation of the magnetic field distribution for the proposed PT lens under point source illumination on the left side, corresponding to Fig. 3.

**Supplementary References**

[S1] D. M. Pozar, *Microwave Engineering, 3rd Edition* (Wiley, 2011).

[S2] L. D. Landau, L. P. Pitaevskii, and E. M. Lifshitz, *Electrodynamics of Continuous Media, Second Edition* (Butterworth-Heinemann, 1984).

[S3] P. B. Johnson and R. W. Christy, "Optical Constants of the Noble Metals," Phys. Rev. B **6**, 4370 (1972).